ICEAI-1080

# A Heuristic for Maximizing the Lifetime of Data Aggregation in Wireless Sensor Networks

**Bing-Hong Liu[a*], Van-Trung Pham[b], Ngoc-Tu Nguyen[a], Yi-Sheng Luo[a]**

[a]Department of Electronic Engineering, National Kaohsiung University of Applied Sciences

415, Chien Kung Rd., Kaohsiung 80778, Taiwan

[b]Department of Information Technology, Pham Van Dong University

99, Hung Vuong Rd., Quang Ngai 57000, Viet Nam

bhliu@kuas.edu.tw

**Abstract**

Recently, many researchers have studied efficiently gathering data in wireless sensor networks to minimize the total energy consumption when a fixed number of data are allowed to be aggregated into one packet. However, minimizing the total energy consumption does not imply the network lifetime is maximized. In this paper, we study the problem of scheduling data aggregation trees working for different time periods to maximize the network lifetime when a fixed number of data are allowed to be aggregated into one packet. In addition, we propose a heuristic to balance the lifetime of nodes in data aggregation trees such that the network lifetime is maximized. Simulation results show that the proposed heuristic provides a good performance.

## 1. Introduction

Wireless sensor networks (WSNs) consist of many wireless sensors, where sensors are often deployed in a wide range of field. Today, many applications of WSNs have been developed, such as surveillance monitoring, seismic monitoring, and fire detection [1], [2], [3]. In these applications of WSNs, one of the most important operations is data gathering. The data-gathering mechanism in WSNs is responsible to collect sensed data from sensors and report the data to a specific node, called the sink. In this paper, we study prolonging the lifetime of WSNs with efficient data gathering.

Tree is a well-known structure that can lead the generated data to the sink in WSNs. With tree structures, each node in the tree is responsible for receiving data from its child nodes and forwarding its generated or received data to its parent node. Recently, many studies investigate designing data gathering trees to maximize the lifetime of WSNs. In [4], the problem of choosing a maximum lifetime tree from a set of shortest path trees is studied. In [5], to solve the maximum lifetime data collection problem in sensor networks, an approximation method that uses intelligent selection of trees is proposed. In [6], when a single sink exists, an approximation algorithm of constructing a spanning tree to prolong the network lifetime is proposed. In the studies [4], [5], [6], the data are directly forwarded through the tree to the sink without aggregating data, and therefore, lots of energy is spent for data gathering.

To minimize the energy consumption for data forwarding, data aggregation is often used to minimize the size of forwarding data [7], [8], [9]. A data aggregation tree is thus used to prolong the network lifetime while data are allowed to be aggregated. In addition, some sensors in WSNs, called source nodes, are responsible for sensing the environment and generating sensing data; and some sensors, called relay nodes, can be used to forward received data to others. When any source node depletes its energy, the end of the network lifetime for the WSN is reached because the network cannot provide adequate quality of service for applications. To prolong the network lifetime, a set of data aggregation trees that work in turn for different time periods can be applied in WSNs. This motivates us to study the problem of scheduling data aggregation trees for different time periods when a fixed number of data are allowed to be aggregated into one packet such that the network lifetime is maximized.

The rest of this paper is organized as follows. Section 2 describes the details of the network model. Section 3 proposes the proposed heuristic. In Section 4, the performance of the proposed heuristic is evaluated in terms of the network lifetime in WSNs. In Section 5, this paper is concluded.

## 2. System Model and Assumptions

In this paper, the communication model in the wireless sensor network is assumed to be a unit disk graph model [10]. In this model, a sensor u can receive messages from sensor v if u is within the transmission range of v. Hereafter, u is said to be v's neighboring node in the network if u can receive messages from v. When all sensors have the same transmission ranges, the WSN can be represented as a connected weighted graph $G(V_G, E_G, w_G, \rho_G)$, where node $v \in V_G$ represents a sensor in the WSN, edge $(u,v) \in E_G$ represents that u and v can communicate with each other, $w_G(v)$ represents the energy of v, and $\rho_G(v)$ represents the number of units of raw data generated by v to report to sink $s \in V_G$ per unit time, where the sink s is a special node in the network and is responsible for data collecting, processing, and analysis. The data generated in a unit time have to be reported to the sink in the same unit time, which is called a working round hereafter. In addition, the nodes v with $\rho_G(v) > 0$ are called source nodes; and the nodes v with $\rho_G(v) = 0$ are called relay nodes. Note that the relay node can receive data from other nodes and forward the received data to the next node for reporting data to the sink. In addition, the source node can generate its own raw data for each working round and works like a relay node to help relaying data.

A data aggregation tree constructed for the network G is a spanning tree T = ($V_T$,$E_T$), where $V_T$ = $V_G$ and $E_T$ ⊆ $E_G$. A data aggregation tree T has to satisfy that T is a connected graph rooted at the sink without cycle. In addition, the data generated by source nodes in G are required to be forwarded to and collected in the sink via T. Because each node u in a data aggregation tree T has to forward its generated or received data to its parent node in T, u has to help forwarding all data generated in the subtree rooted at u. Let u.tot denote the total number of units of raw data generated and received by u. Also let $u_T$.CH denote the set of u's child nodes in T. Then u.tot can be calculated by the following equation:

$$u.tot = \rho_G(u) \sum_{v \in u_T.CH} v.tot \qquad (1)$$

Assume that raw data generated in the WSN can be aggregated in each node in T. Let $\alpha \in Z^+$ ($\alpha \geq 2$) be the aggregation ratio, representing the maximum number of units of raw data allowed to be aggregated into one unit-size packet [7], [8], [9]. The number of unit-size packets that are forwarded by each node u in T, denoted by u.δ, can be calculated by the following equation:

$$u.\delta = \left\lceil \frac{u.tot}{a} \right\rceil \qquad (2)$$

Let $e_{tx}$ (or, $e_{rx}$) represent the energy consumption of a sensor node to transmit (or, receive) one unit-size packet. Because each node u in T has to receive packets from its child nodes in T and forward the aggregated data to its parent node in T, the energy consumption for node u in T to receive and forward data within each working round is therefore calculated by the following equation:

$$u.eng = e_{rx} \sum_{v \in u_T.CH} v.\delta + e_{tx} u.\delta \qquad (3)$$

In WSN, each node u has limited energy $w_G(u)$ to maintain its activity. The lifetime of u, represented by u.ℓ, is therefore defined by the maximum number of working rounds for u to sustain [11], [12], [13], calculated by the following equation:

$$u.\ell = \left\lfloor \frac{w_G(u)}{u.eng} \right\rfloor \qquad (4)$$

In a data aggregation tree T, every node has to help forwarding the aggregated data to its parent node in T. Therefore, the lifetime of T, denoted by T.ℓ, is defined as the minimum u.ℓ for all u ∈ T, as calculated by the following equation:

$$T.\ell = \min(v_1.\ell, v_2.\ell, \ldots, v_k.\ell) \qquad (5)$$

where $v_1, v_2, \ldots, v_k \in$ T. Because a set of data aggregation trees, $T_1, T_2, \ldots, T_p$, can work in turn for time periods $t_1, t_2, \ldots, t_p$, respectively, the data aggregation trees can be employed to prolong the network lifetime. The network lifetime is thus calculated by the sum of the lifetime of all scheduled data aggregation trees, that is, $\sum_{i=1}^{p} T_i.\ell$. Therefore, our problem is to schedule data aggregation trees for different time periods when a fixed number of data are allowed to be aggregated into one packet such that the network lifetime is maximized.

## 3. Proposed Heuristic

In this paper, a heuristic is proposed for the problem of scheduling data aggregation trees for different time periods such that the network lifetime is maximized. The idea of the proposed heuristic is to find a data aggregation tree with maximum lifetime for any WSN when a new data aggregation tree is required for the network. In the proposed heuristic, a shortest path tree T is first constructed to span all nodes in the WSN because the path from any node in the tree to the sink is the shortest. For the purpose, while given a wireless sensor network G, a breadth-first-search method [14] can be used to construct a shortest path tree T to span all nodes in G.

After constructing a shortest path tree T in the network G, our idea is to rebuild the tree structure T such that the nodes having little or no energy become leaf nodes. This is because a leaf node in a tree does not need to relay other nodes' data, that is, the leaf node wastes energy only on transmitting its generated data. When a non-leaf node in T is changed to a leaf node, the node can save its energy on forwarding data.

After transferring the non-leaf nodes having little or no energy into leaf nodes in T, we then check every node u in T to see if a local adjustment can be made to prolong the lifetime of u's parent node in T. Let u's parent node be x. Also let v be the neighboring node of u in the network G such that $v.h_T = u.h_T - 1$ and v is not u's parent node in T, where $v.h_T$ (or $u.h_T$) denotes the minimum hop count from the sink to v (or u) via the tree structure T. We check that if the parent node of u is changed from x to v and the lifetime of u's parent node in T is therefore prolonged, edge (u,v) is inserted into T and edge (u,x) is deleted from T.

## 4. Performance Evaluation

In order to evaluate the performance of the proposed heuristic, the simulations are performed by using DevC++ modeler environment. In this section, the scenario and the parameters used for the simulations are presented in Section 4-A. In addition, the results of performance evaluation of the proposed heuristic are presented in Section 4-B.

### A. Scenario and the Parameters

Here, simulations were conducted to evaluate the performance of the proposed heuristic. In the simulation, the WSNs were generated by randomly deploying sensors when the number of nodes ranging from 200 to 1000 including source nodes and 50 relay nodes deployed in a 10 × 10 field. In addition, the transmission range and the initial energy of each sensor was set to 2 and $10^5$, respectively. Because the transmission power is about double the reception power [8], $e_{tx}$ and $e_{rx}$ were set to 2 and 1, respectively. The numbers of units of raw data generated by source nodes per unit time were randomly selected from the interval [1,10]. In the following simulation, 100 WSNs were generated. In addition, the simulation results were obtained by averaging the data of 100 networks. In the simulation, a shortest-path-tree-based scheduling algorithm (SPTBSA) was introduced here to compare with the proposed heuristic in terms of the network lifetime.

**B. Network Lifetime**

Fig. 1 shows the simulation results for the network lifetime versus the number of network nodes. It is clear that when the number of network nodes increases, the network lifetime of the proposed heuristic and the SPTBSA decreases. This is because more source nodes exist in the networks, and more data must be relayed to the sink. Therefore, more energy must be consumed in some critical nodes. In addition, our method have a longer network lifetime than that of the SPTBSA. This is because the lifetime of critical nodes is prolonged our proposed heuristic.

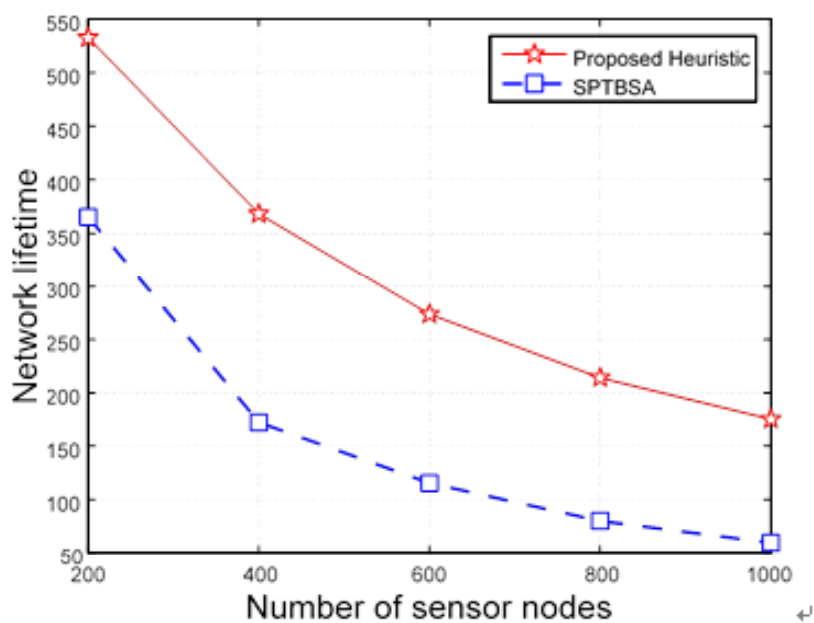

Figure 1. The network lifetime versus the number of network nodes ranging from 200 to 1000.

**5. Conclusion**

In this paper, we study the problem of scheduling data aggregation trees for different time periods when a fixed number of data are allowed to be aggregated into one packet such that the network lifetime is maximized. In addition, we propose a heuristic for the problem. In the proposed heuristic, we find a data aggregation tree with maximum lifetime for any WSN when a new data aggregation tree is required for the network. While finding a data aggregation tree, a shortest path tree is constructed to be the data aggregation tree to span all nodes in the WSN. Then, the tree structure is rebuilt such that the lifetime of the data aggregation tree can be maximized. The simulation results show that our proposed heuristic provides a good performance.

## Acknowledgment

This work was supported by the Ministry of Science and Technology under Grant MOST 104-2221-E-151-014.